\documentclass[preprint2]{aastex61}
\usepackage{natbib,graphicx,longtable,afterpage,amsmath}

\shorttitle{Evolution of the $M_{\rm BH}-M_{\rm stellar}$ Relation up to $z\sim2.5$}
\shortauthors{Suh et al.}

\def\Ha{\text{H$\alpha$}}
\def\Hb{\text{H$\beta$}}
\def\MgII{\text{Mg\,\textsc{ii}}}
\def\FeII{\text{Fe\,\textsc{ii}}}
\def\NII{\text{[N\,\textsc{ii}}]}

\def\OIII{\text{[O\,\textsc{iii}}]}

\begin{document}

\title{No significant Evolution of relations between black hole mass and galaxy total stellar mass up to $z\sim2.5$}

\author{Hyewon Suh}
\altaffiliation{Subaru Fellow}
\email{suh@naoj.org}
\affil{Subaru Telescope, National Astronomical Observatory of Japan (NAOJ), 650 North A'ohoku place, Hilo, HI 96720, USA}

\author{Francesca Civano}
\affiliation{Harvard-Smithsonian Center for Astrophysics, Cambridge, MA, 02138, USA}

\author{Benny Trakhtenbrot}
\affiliation{School of Physics and Astronomy, Tel Aviv University, Tel Aviv 69978, Israel}

\author{Francesco Shankar}
\affiliation{Department of Physics and Astronomy, University of Southampton, Highfield SO17 1BJ, UK}

\author{G\"unther Hasinger}
\affiliation{European Space Astronomy Centre (ESA/ESAC), Director of Science, E-28691 Villanueva de la Ca\~{n}ada, Madrid, Spain}

\author{David B. Sanders}
\affiliation{Institute for Astronomy, University of Hawaii, 2680 Woodlawn Drive, Honolulu, HI 96822, USA}

\author{Viola Allevato}
\affiliation{Scuola Normale Superiore, Piazza dei Cavalieri 7, I-56126 Pisa, Italy}

\begin{abstract}
We investigate the cosmic evolution of the ratio between black hole mass ($M_{\rm BH}$) and host galaxy total stellar mass ($M_{\rm stellar}$) out to $z\sim2.5$ for a sample of 100 X-ray-selected moderate-luminosity, broad-line active galactic nuclei (AGNs) in the {\it Chandra}-COSMOS Legacy Survey. By taking advantage of the deep multi-wavelength photometry and spectroscopy in the COSMOS field, we measure in a uniform way the galaxy total stellar mass using a SED decomposition technique and the black hole mass based on broad emission line measurements and single-epoch virial estimates. Our sample of AGN host galaxies has total stellar masses of $10^{10-12}M_{\odot}$, and black hole masses of $10^{7.0-9.5}M_{\odot}$. Combining our sample with the relatively bright AGN samples from the literature, we find no significant evolution of the $M_{\rm BH}-M_{\rm stellar}$ relation with black hole-to-host total stellar mass ratio of $M_{\rm BH}/M_{\rm stellar}\sim0.3\%$ at all redshifts probed. We conclude that the average black hole-to-host stellar mass ratio appears to be consistent with the local value within the uncertainties, suggesting a lack of evolution of the $M_{\rm BH}-M_{\rm stellar}$ relation up to $z\sim2.5$.
\end{abstract}

\section{Introduction}

The local universe provides clear evidence that the growth of supermassive black holes (SMBHs) is closely connected with galaxy evolution, as revealed by well-known tight correlations between the black hole (BH) mass and their host bulge properties (i.e., $M_{\rm BH}-M_{\rm bulge}$, $M_{\rm BH}-\sigma$ relations; \citealt{Magorrian98, Ferrarese00, Gebhardt00, Haring04, Gultekin09, Sani11, McConnell13, Kormendy13, Shankar16}). Several studies have reported that the mass ratio between BH and host bulge is $M_{\rm BH}/M_{\rm bulge}\sim10^{-3}$ with intrinsic dispersion of $\sim0.3$ dex (e.g., \citealt{Marconi03, Haring04, Sani11}). \citet{Kormendy13} argued for larger values of $M_{\rm BH}/M_{\rm bulge}\sim10^{-2.3}$, mainly because of their revision of increased BH masses. Many theoretical models have proposed active galactic nucleus (AGN) feedback to explain these physical connection (e.g., \citealt{Silk98, DiMatteo05, Hopkins06}), and yet, how SMBHs and their host galaxies evolve onto these local scaling relations through cosmic time remains unclear. 

While relations between BH mass and bulge properties in quiescent bulge-dominant galaxies show the tightest correlation, for high redshift studies ($z>1$) it is difficult to estimate the bulge mass due to the lack of spatial resolution and sensitivity, and thus, the total stellar mass is used instead, estimated by assuming a mass-to-light ratio or a spectral energy distribution (SED) fitting (e.g., \citealt{Merloni10, Cisternas11, Bongiorno14}). Measuring the BH mass is also challenging beyond the local universe, and the virial method is often used to estimate the BH mass for galaxies hosting broad-line AGN (e.g., \citealt{Kaspi00, Vestergaard02, Woo02, McLure02, McLure04, Greene05, Kollmeier06, Vestergaard06, Shen08, Shen11, Trakhtenbrot12}). 

\citet{Reines15} have quantified the relationship between BH mass and total stellar mass for nearby galaxies to facilitate work at higher redshifts, including both galaxies with quiescent and active BHs. They found that local AGN host galaxies tend to fall below the canonical BH-to-bulge mass relations defined by inactive early-type galaxies at a given total stellar mass, by more than an order of magnitude, with $M_{\rm BH}/M_{\rm stellar}\sim10^{-4.6}$. \citet[see also \citealt{Bernardi07, Shankar17}]{Shankar16} used detailed Monte Carlo simulations to put forward evidence for a possible bias in the local $M_{\rm BH}-M_{\rm stellar}$ relation of inactive BHs with dynamically-measured masses. They showed that, especially in early-type galaxies, the necessary requirement of resolving the gravitational sphere of influence of the central BH for reliable dynamical mass measurements can, by itself, bias upwards the $M_{\rm BH}-M_{\rm stellar}$ relation, and proposed an intrinsic scaling $M_{\rm BH}-M_{\rm stellar}$ relation, in the hypothesis that the mass of the BH is predominantly dependent on velocity dispersion. \citet{Shankar19} further showed that local AGNs, which do not suffer from this selection effect, tend instead to naturally sit around the intrinsic $M_{\rm BH}-M_{\rm stellar}$  relation proposed by \citet[see also \citealt{Reines15}]{Shankar16}.

Many observational studies have found that SMBHs beyond the local universe are over-massive at a given host stellar mass compared with that at the present time, suggesting that BHs were able to grow more efficiently than their host galaxies (e.g., \citealt{Peng06, Treu07, Woo08, Merloni10, Decarli10, Trakhtenbrot10, Bennert11, Caplar15, Park15, Trakhtenbrot15, Caplar18}). \citet{Peng06} suggested that the $M_{\rm BH}/M_{\rm stellar}$ ratio is by a factor of $>\sim4$ times larger at $z>1.7$ than today. \citet{Decarli10} also claimed that the BH-to-host mass ratios significantly increase by a factor of $\sim7$ at $z=3$ from a sample of 96 quasars. \citet{Merloni10} reported that the average BH to host galaxy mass ratio appears to evolve positively with redshift, with $M_{\rm BH}/M_{\rm stellar}\propto(1+z)^{0.68}$. 

On the other hand, some others found that the relationship between SMBHs and their host masses matches the correlation that we observe today, suggesting no evolution in the scaling relation within the uncertainties (e.g., \citealt{Shields03, Salviander07, Shen08, Shankar09, Cisternas11, Schramm13, Salviander15, Shen15, Sun15}). \citet{Jahnke09} suggested no evolution in the $M_{\rm BH}-M_{\rm stellar}$ relation using 10 of the targets in \citet{Merloni10} sample when they independently derive host galaxy properties using {\it HST} observations. They found some evidence of substantial disk components from their {\it HST} imaging, suggesting that if the objects were purely bulge-dominated, the $M_{\rm BH}-M_{\rm stellar}$ relation has not evolved, or at least not as rapidly as the relations between BH mass and spheroid properties since $z\sim2$. 

However, selection biases could affect the interpretation of the results on these relations at high redshift (e.g., \citealt{Lauer07, Treu07, Schulze14, Shen15}). \citet{Lauer07} pointed out that high redshift samples are generally selected by nuclear activity (i.e., AGN luminosity), and therefore, biased toward the most luminous AGNs. Such luminous systems are intrinsically rare and do not represent the typical AGN population \citep{Richards06, Ross13, Aird15}. Due to this bias, luminous AGNs are more likely to be found in less massive galaxies at higher redshift, resulting in an apparent evolution of the $M_{\rm BH}-M_{\rm stellar}$ relation.

In this paper, we make use of a large, deep/uniform X-ray depth and the extensive multi-wavelength photometric and spectroscopic data of {\it Chandra}-COSMOS Legacy Survey \citep{Civano16} to investigate the cosmic evolution of the relationship between BH mass and galaxy total stellar mass up to $z\sim2.5$. The redshift range of $1<z<3$ is an essential cosmic period for studying the link between BH growth and galaxy evolution, which corresponds to the peak epoch of star formation and AGN activity. Our sample of X-ray-selected AGNs with lower luminosities ($L_{\rm bol}\sim10^{44-46}~{\rm erg~s^{-1}}$), together with the deep high-quality spectroscopy, contains a much more representative population of SMBHs and host galaxies, and is therefore less susceptible to the selection biases induced by flux limit effects, as compared to previous, mostly optical, samples. The BH mass estimate is based on the single-epoch virial method using the Keck/DEIMOS optical and Subaru/FMOS NIR spectroscopy, and the stellar mass is measured via a multi-component spectral energy distribution (SED) fitting. These measurements then allow us to investigate the evolution of BH-galaxy scaling relations up to $z\sim2.5$, for the deepest/largest data set adopted so far in this kind of studies. 

Throughout this work, we adopt a standard $\Lambda$CDM cosmology with $\Omega_{m}=0.3,~\Omega_{\Lambda}=0.7$, and $H_{0}=70~{\rm km~s^{-1}~Mpc^{-1}}$.

\section{Broad-line AGN sample and data}
  
\begin{figure*}
\centering
\includegraphics[width=1\textwidth]{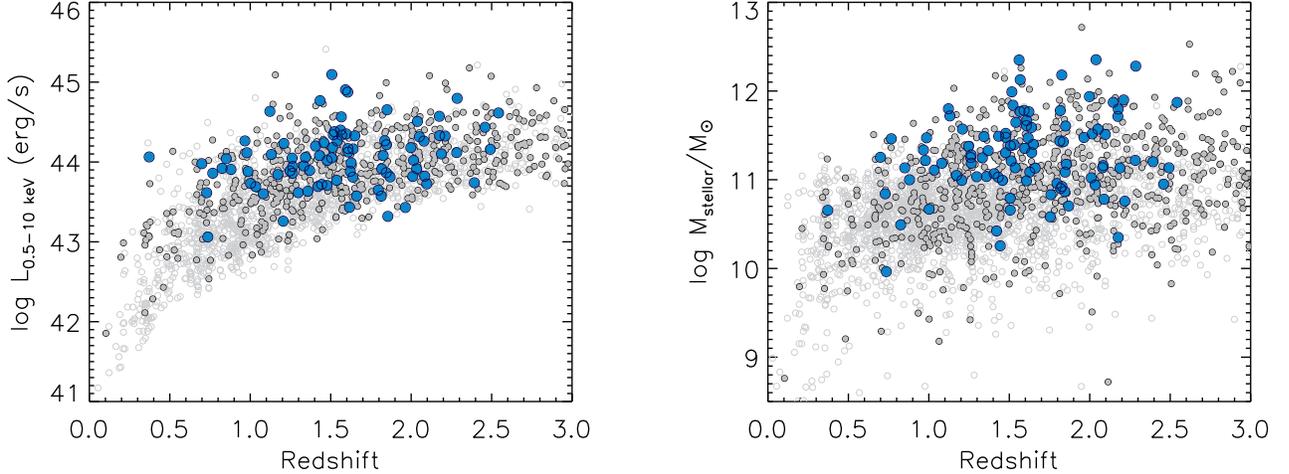}
\caption{Absorption-corrected 0.5--10 keV X-ray luminosity (left) and stellar mass (right) vs. redshift (spectroscopic or photometric) distribution for all X-ray sources in the CCLS. The sources classified as Type 1 (optically unobscured and/or broad-line) AGNs are denoted by filled gray circles (see \citealt{Marchesi16, Suh19} for details). Our final sample of broad-line AGNs is shown as filled blue circles.}
\label{fig:zdist}
\end{figure*}

We construct our sample of broad-line AGNs from the {\it Chandra}-COSMOS Legacy Survey (CCLS; \citealt{Civano16}) with a multi-wavelength data set \citep{Marchesi16, Laigle16}, which contains a total of 4016 X-ray sources down to a flux limit of $2.2\times10^{-16}$, $1.5\times10^{-15}$, and ${\rm 8.9\times10^{-16}~erg~cm^{-2}~s^{-1}}$ at 20 per cent completeness in the $0.5-2$ keV, $2-10$ keV, and $0.5-10$ keV bands. Recently, \citet{Hasinger18} presented the first comprehensive spectroscopic observations with Keck/DEIMOS \citep{Faber03} over a large area of $\sim2.2~\rm deg^{2}$ in the COSMOS field (DEIMOS 10K spectroscopic survey). Compared with the previous spectroscopic surveys in the COSMOS field, such as the zCOSMOS bright spectroscopic catalog \citep{Lilly07} containing 10,644 spectra for a sample of galaxies with $I_{\rm AB}<22.5$ mag, the DEIMOS 10K spectroscopic catalog provides spectroscopy with a spectral resolution $R\sim2000-2700$ for a sample of 10,718 objects with $I_{\rm AB}<23.5-25$ mag, including the newly-detected deep {\it Chandra} X-ray sources selected from CCLS. With the deeper magnitude limits, we expect to obtain a lower limit to the BH masses of $\log M_{\rm BH}\sim7.04 M_{\odot}$ at $z=2$ using the methods detailed in Section~\ref{sec:Mbh}. Furthermore, the FMOS-COSMOS spectroscopic survey was conducted with the Subaru/FMOS \citep{Kimura10} NIR high-resolution spectrographs ($R\sim2200$), described in detail in \citet{Silverman15} (see also \citealt{Kashino13, Schulze18}).  Our sample, therefore, represents the most typical AGN population covering lower luminosities, compared to previous studies at $z>1$. 

We select our sample of broad-line AGNs, for which one or more broad emission lines with a full width at half-maximum (FWHM) larger than 2000 ${\rm km~s^{-1}}$ have been identified by analyzing optical spectra of 1078 sources with DEIMOS, and additionally NIR spectra of 589 sources with FMOS. We detect 21~\Ha,~5~\Hb,~and 74~\MgII~broad emission lines, for which having S/N greater than 10 per pixel, in the optical/NIR spectra, respectively. We finally obtain 100 broad-line AGNs, covering the redshift range $z=0-2.5$ (Table~\ref{tbl:sample}).

In Figure~\ref{fig:zdist}, we show the distribution of the absorption-corrected 0.5--10 keV X-ray luminosity (left) and total stellar mass (right; \citealt{Suh19}) as a function of redshift for all the CCLS sources. The final sample of our broad-line AGNs covers the full region of the X-ray luminosity ($L_{\rm 0.5-10~keV}\sim10^{43-45}~{\rm erg~s^{-1}}$) and stellar mass ($M_{\rm stellar}\sim10^{10-12}M_{\odot}$) as the overall CCLS Type 1 AGN sample spanning about 2 dex at $1<z<2$. 

\section{Host Stellar Mass}\label{sec:Ms}

The total stellar mass of our sample of broad-line AGN host galaxies is adopted from \citet{Suh19} by performing a multi-component SED fitting from far-IR (500$\mu$m) to near-UV (2300\AA). \citet{Suh19} decomposed the SED using an AGN accretion disk emission model (i.e., big blue bump) from \citet{Richards06}, a dust torus model from \citet{Silva04}, a starburst template from \citet{Chary01} and \citet{Dale02}, and a galaxy stellar population model of \citet{BC03}. For the galaxy model template, they used the \citet{Chabrier03} initial mass function (IMF) and an exponentially decaying star formation histories with characteristic times ranging from $\tau$=0.1 to 30 Gyr and constant star formation. The law of \citet{Prevot84} for an AGN accretion disk template and the law of \citet{Calzetti00} for a set of galaxy templates are used to take into account the reddening effect. Four dust torus templates are used depending on the amount of nuclear obscuration.  

Due to the degeneracy between the AGN accretion disk and the galaxy emission in the UV-optical wavelengths of SED fitting, we further require that the AGN emission dominates the galaxy light in the UV bands ($>50$\% of total) by the fact that the UV/optical emission should come from the AGN accretion disk for Type 1 AGNs. Therefore, the stellar mass for the best-fit with this constraint could be biased toward the upper limit (see \citealt{Suh19}). The typical uncertainties for the stellar mass are $+0.19$ dex and $-0.36$ dex toward higher and lower masses, respectively. A full detailed description is presented in \citet*{Suh17, Suh19}.

\section{BH mass and Accretion rate}\label{sec:Mbh}

\begin{figure*}
\centering
\includegraphics[width=0.49\textwidth]{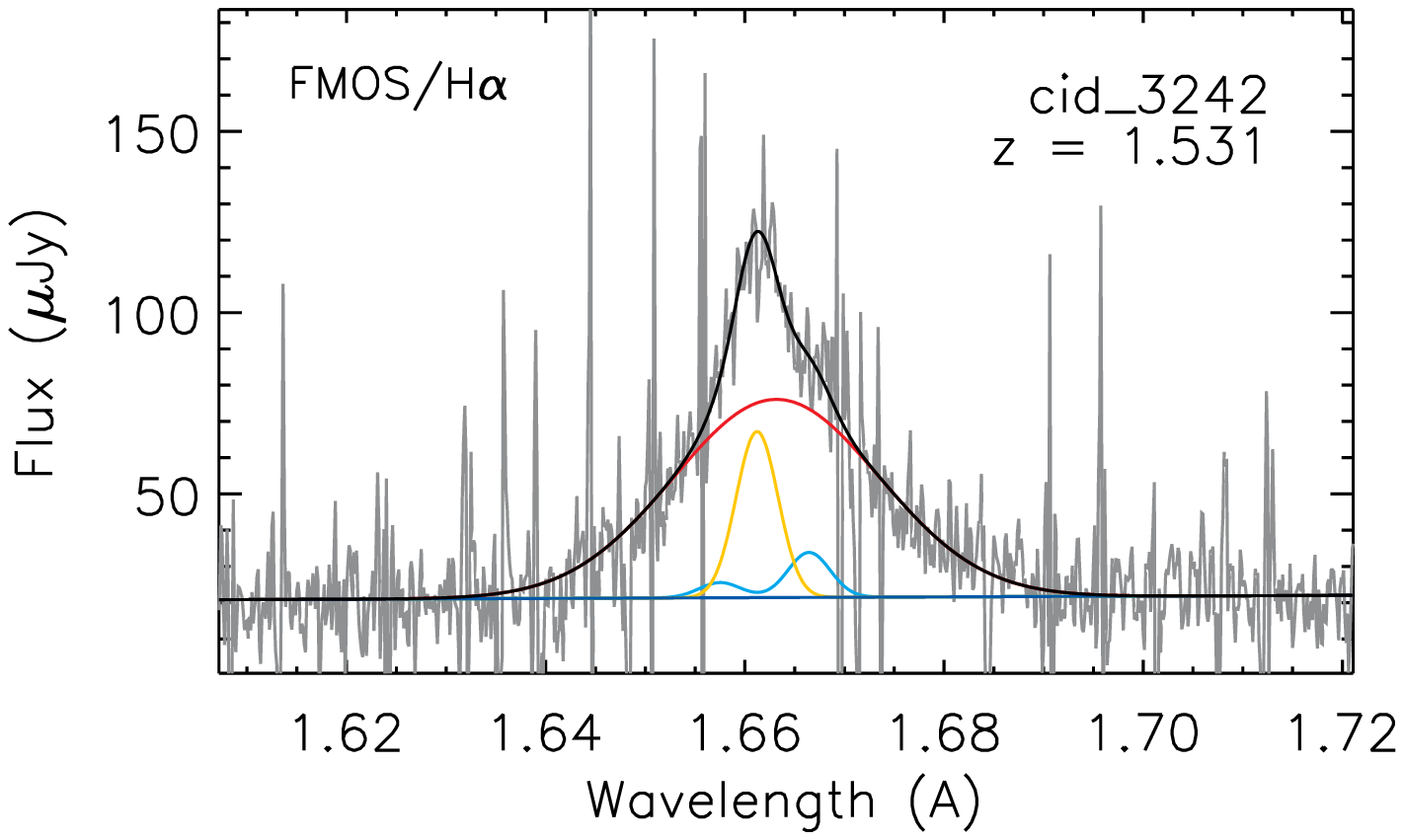}
\includegraphics[width=0.49\textwidth]{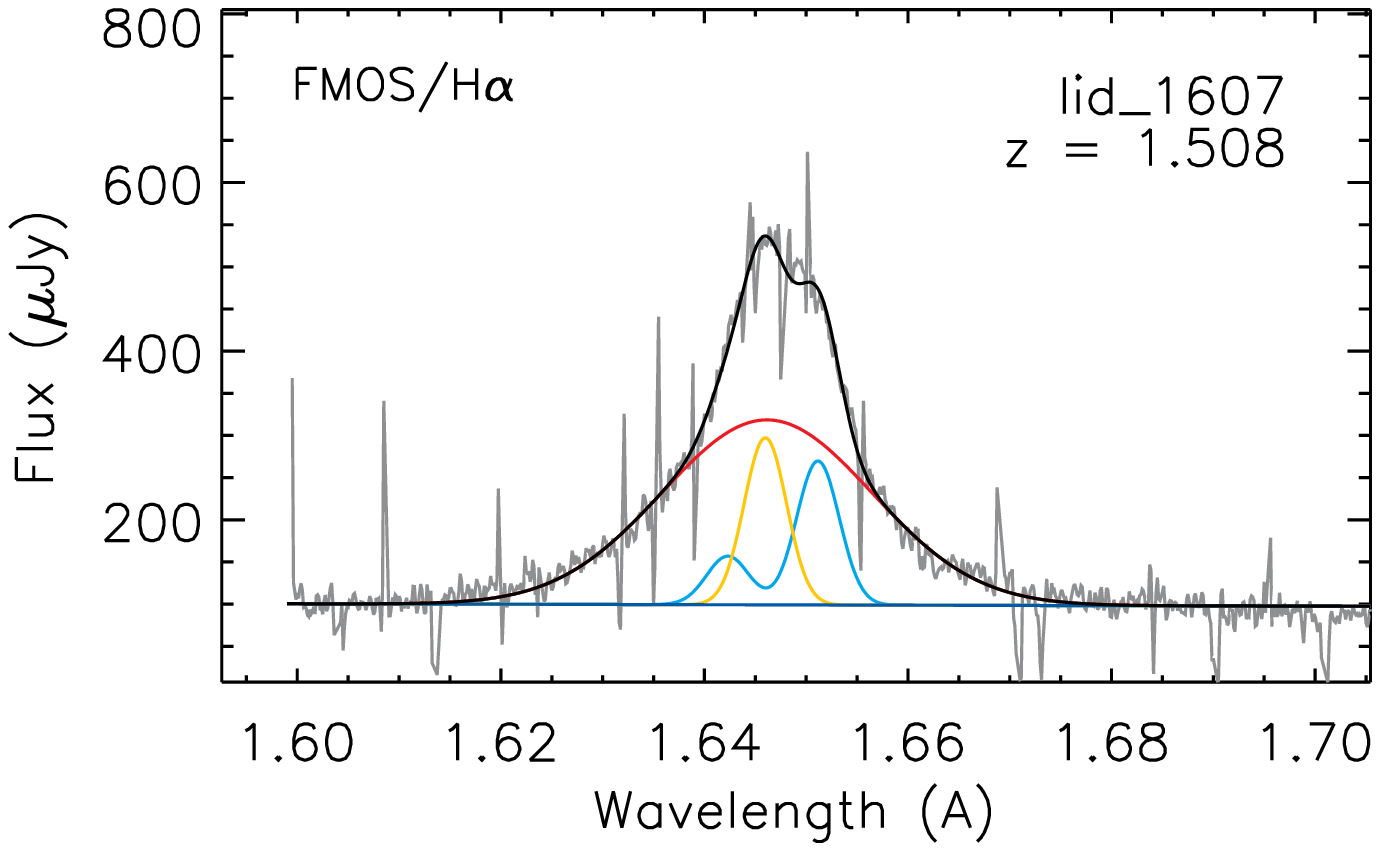}
\includegraphics[width=0.49\textwidth]{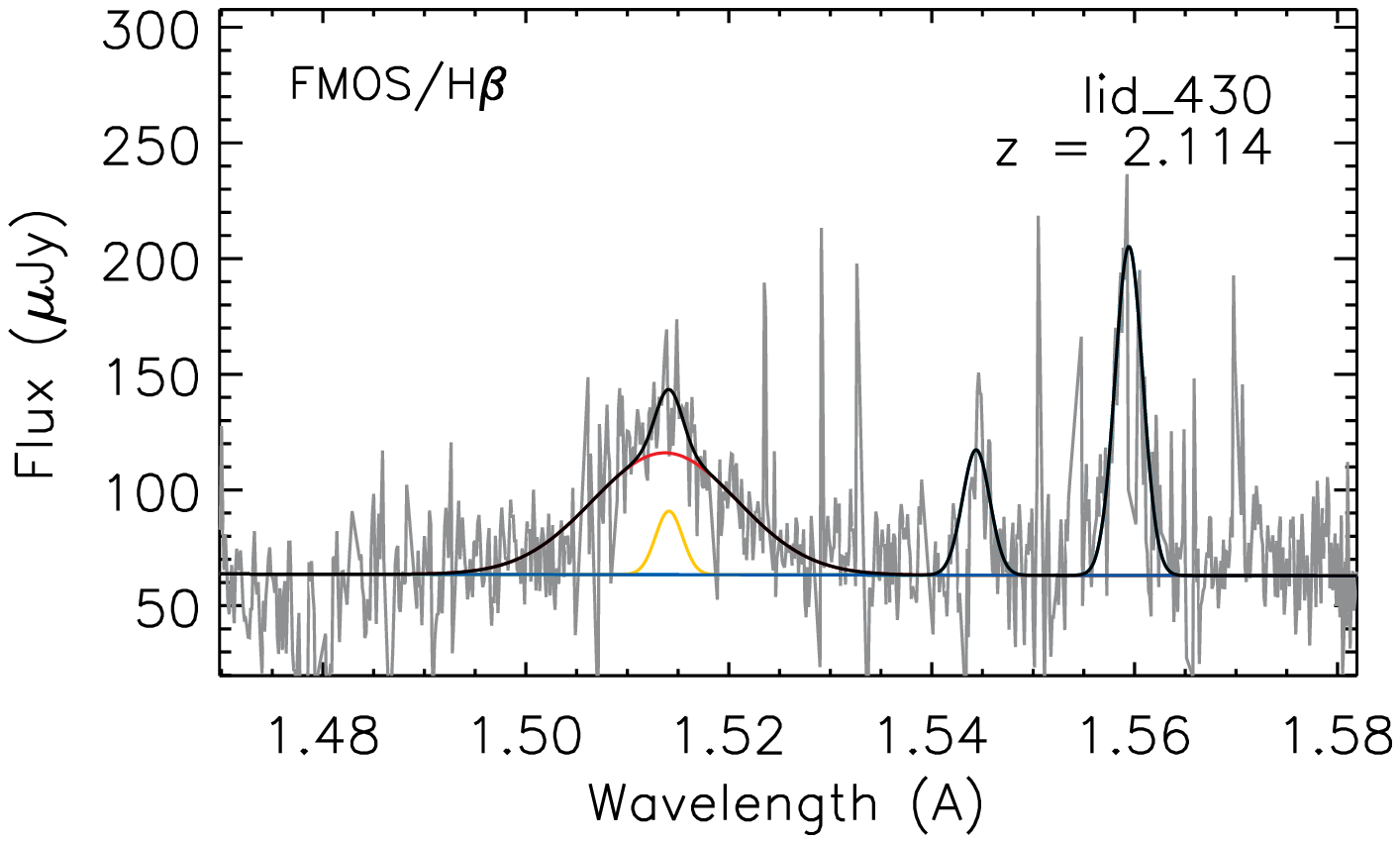}
\includegraphics[width=0.49\textwidth]{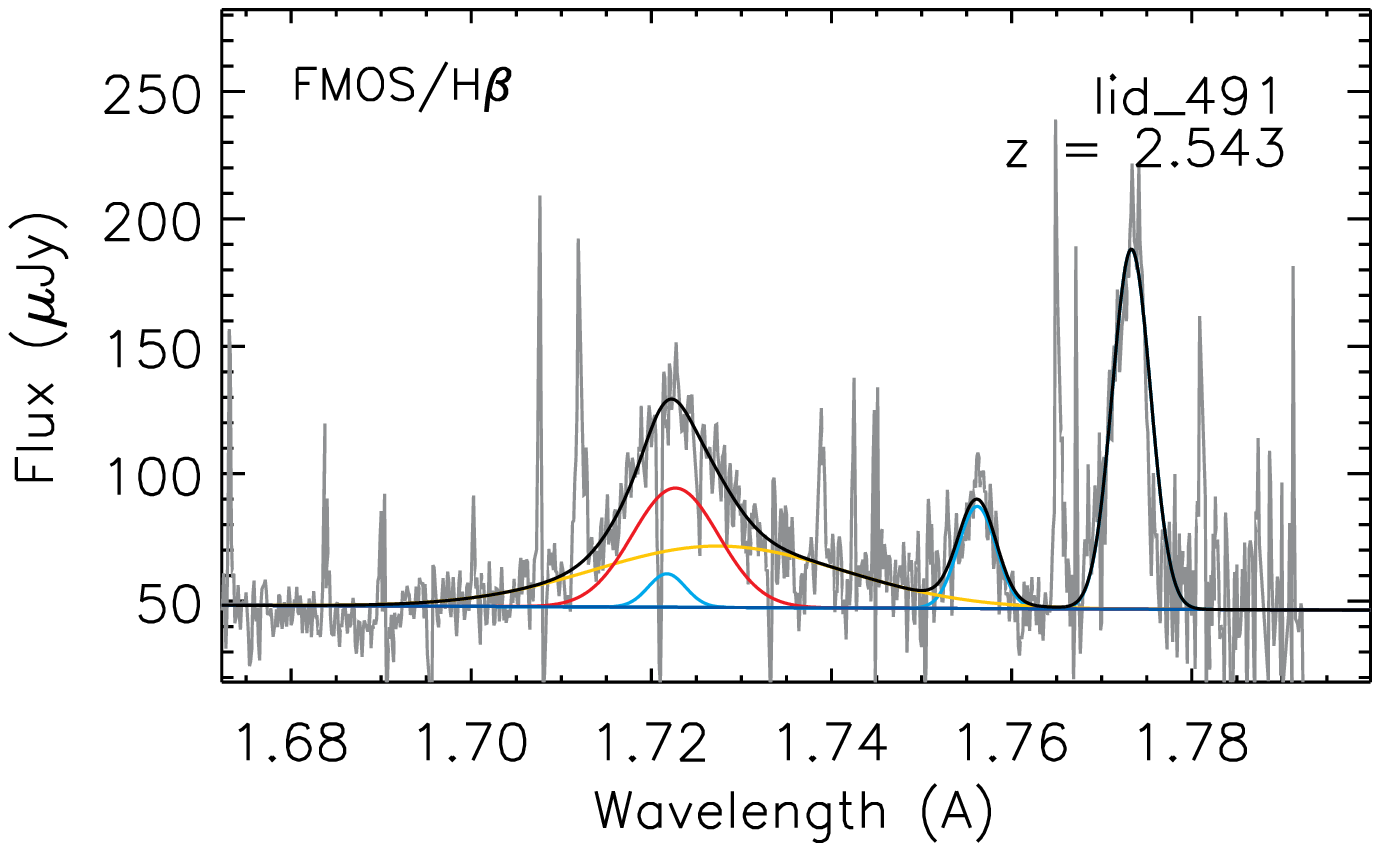}
\includegraphics[width=0.49\textwidth]{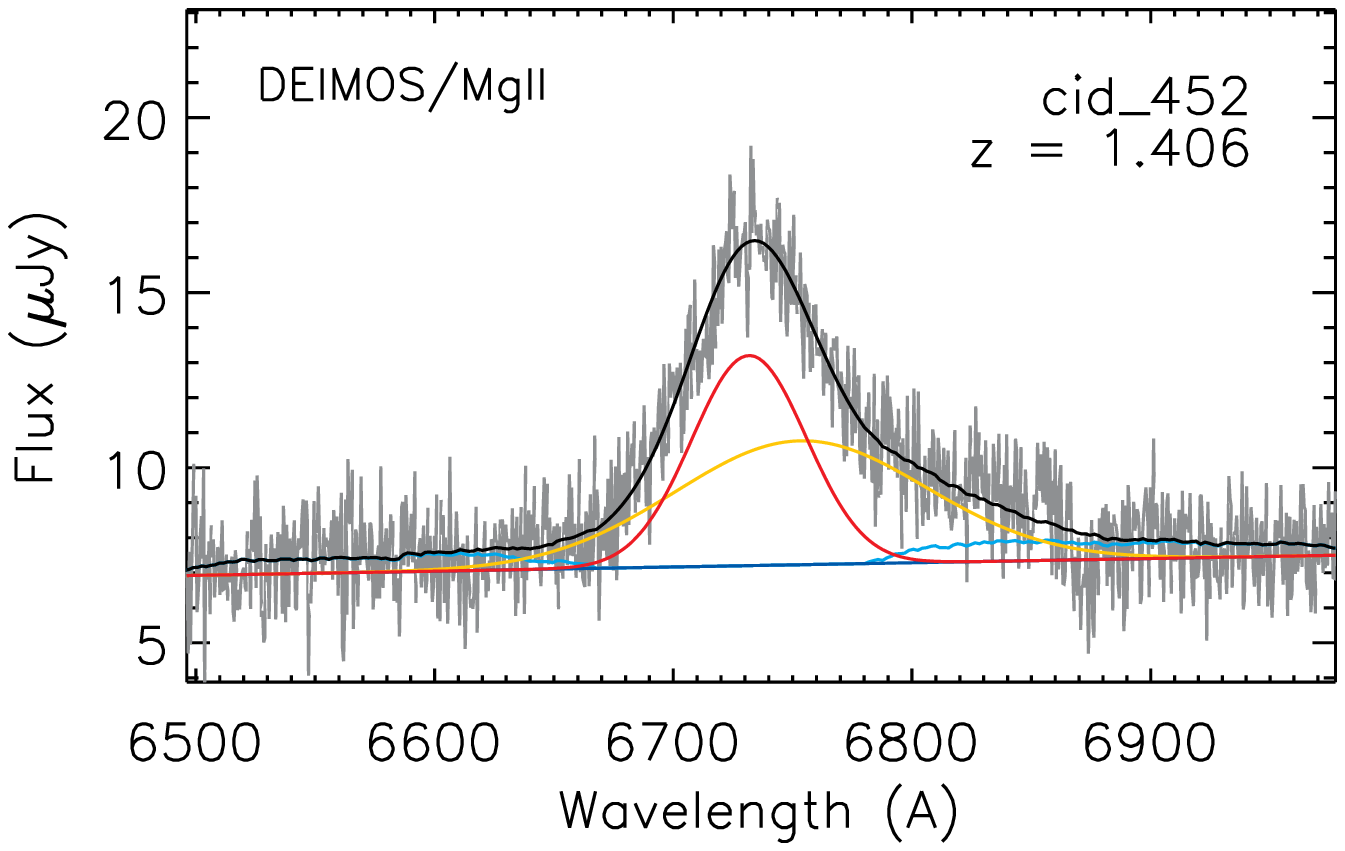}
\includegraphics[width=0.49\textwidth]{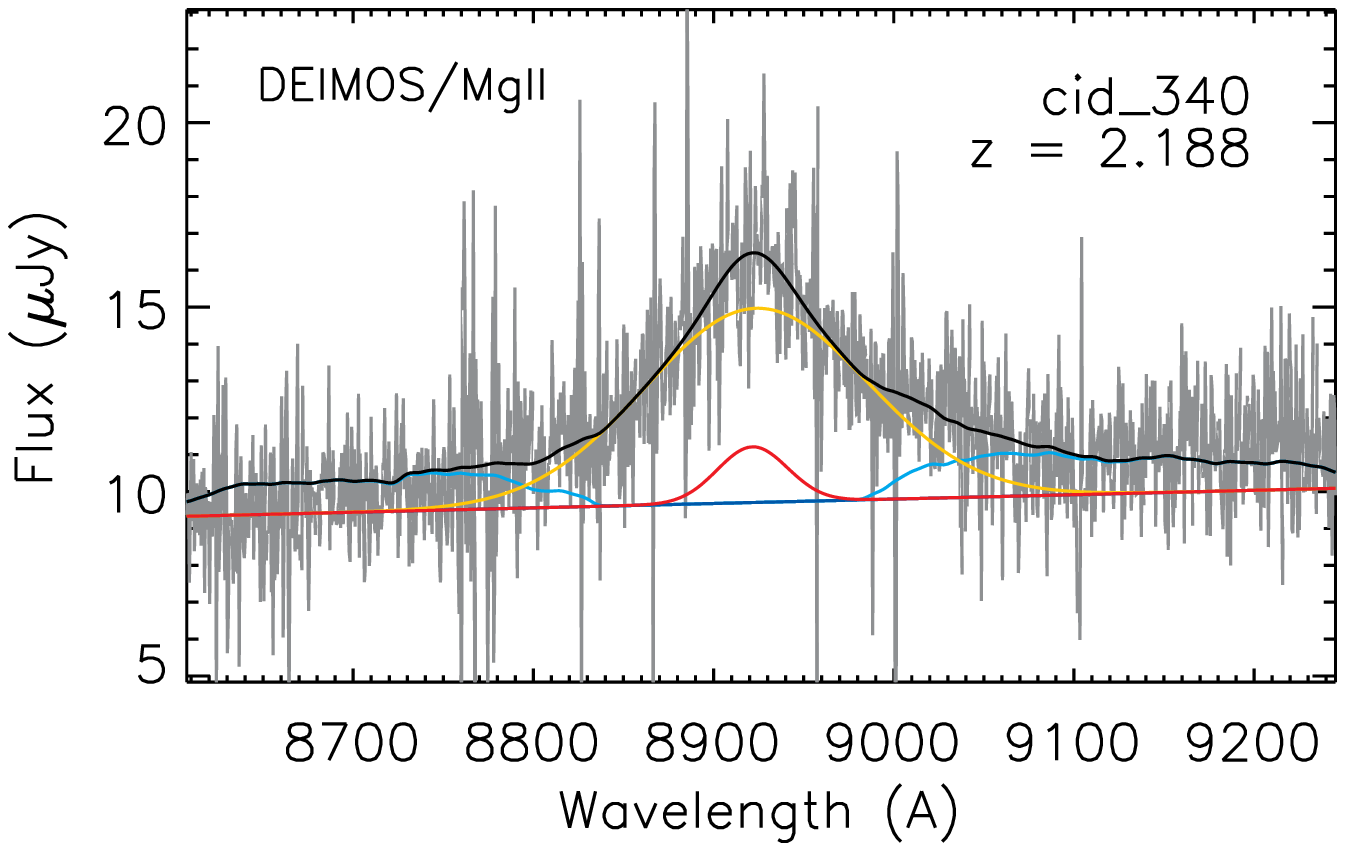}
\caption{Examples of broad-line fits for \Ha~(top panels), \Hb~(middle panels), and \MgII~(bottom panels) emission lines for our sample of AGNs. The observed spectrum (gray) with the best-fit model (black) is indicated. The power-law continuum, narrow-line components (blue), \FeII~emission component (blue), and broad-line components (red and yellow) are also indicated with colored curves.}
\label{fig:Mbh_fits}
\end{figure*}

The virial BH mass can be estimated using the broad-line width and the continuum/line luminosity from the single-epoch, rest-frame UV/optical spectra as a proxy for the characteristic velocity and the size of the broad-line region (BLR; e.g., \citealt{Vestergaard02, McLure02, McLure04, Greene05, Vestergaard06, Shen11, Trakhtenbrot12, Mejia16}). We measure the spectral complexes of broad \Ha,~\Hb,~and~\MgII~emission lines present in optical and NIR spectra, depending on the redshift, to derive the virial BH mass of broad-line AGNs. We use the \texttt{mpfit} routine to fit the emission lines, which adopts a Levenberg-Marquardt least-squares minimization algorithm to derive the best-fit parameters and a measure of the goodness of the overall fit. The full details of the fit procedure are described in detail in \citet{Suh15}. 

\begin{deluxetable*}{rcCCCcc}
\tablecaption{Broad-line AGN Sample \label{tbl:sample}}
\tablecolumns{8}
\tablenum{1}
\tablewidth{0pt}
\tablehead{
\colhead{Object ID} & \colhead{Redshift} &
\colhead{$\log~M_{\rm BH}$} & \colhead{$\log~L_{\rm bol}$} & \colhead{$\log M_{\rm stellar}$} & \colhead{Instrument} & \colhead{Line} \\
\colhead{} & \colhead{} & \colhead{($M_{\odot}$)} & \colhead{(erg s$^{-1}$)} & \colhead{($M_{\odot}$)} &\colhead{} & \colhead{} 
}
\colnumbers
\startdata
    cid-36 & 1.826 & 9.38$\pm$0.06 & 45.63 & 12.18$^{+  0.00} _{-0.04}$ & DEIMOS& \MgII \\
    cid-61 & 1.478 & 8.62$\pm$0.00 & 45.38 & 11.48$^{+  0.00} _{-0.15}$ & DEIMOS& \MgII \\
    cid-66 & 1.512 & 8.45$\pm$0.03 & 45.77 & 11.21$^{+  0.24} _{-0.01}$ & DEIMOS& \MgII \\
    cid-69 & 0.979 & 8.42$\pm$0.08 & 45.68 & 11.22$^{+  0.19} _{-0.06}$ & DEIMOS& \MgII \\
    cid-70 & 1.638 & 8.85$\pm$0.11 & 45.28 & 11.59$^{+  0.16} _{-0.03}$ & DEIMOS& \MgII \\
    cid-87 & 1.606 & 8.77$\pm$0.11 & 46.69 & 11.61$^{+  0.09} _{-0.00}$ & DEIMOS& \MgII \\
    cid-98 & 1.506 & 7.74$\pm$0.05 & 45.59 & 10.79$^{+  0.40} _{-0.64}$ & DEIMOS& \MgII \\
   cid-102 & 1.847 & 8.75$\pm$0.23 & 45.81 & 11.09$^{+  0.14} _{-0.28}$ & DEIMOS& \MgII \\
   ... & ... & ... & ... & ... & ... & ... \\
\enddata
\tablecomments{Column (1) object ID \citep{Marchesi16}. Column (2) redshift. Column (3) BH mass derived from virial method. Column (4) AGN bolometric luminosity. Column (5) total stellar mass derived from SED fitting \citep{Suh19}. Column (6) instrument for spectroscopy. Column (7) broad emission line used for the BH mass measurement.}
\end{deluxetable*}

\begin{figure*}
\centering
\includegraphics[width=0.8\textwidth]{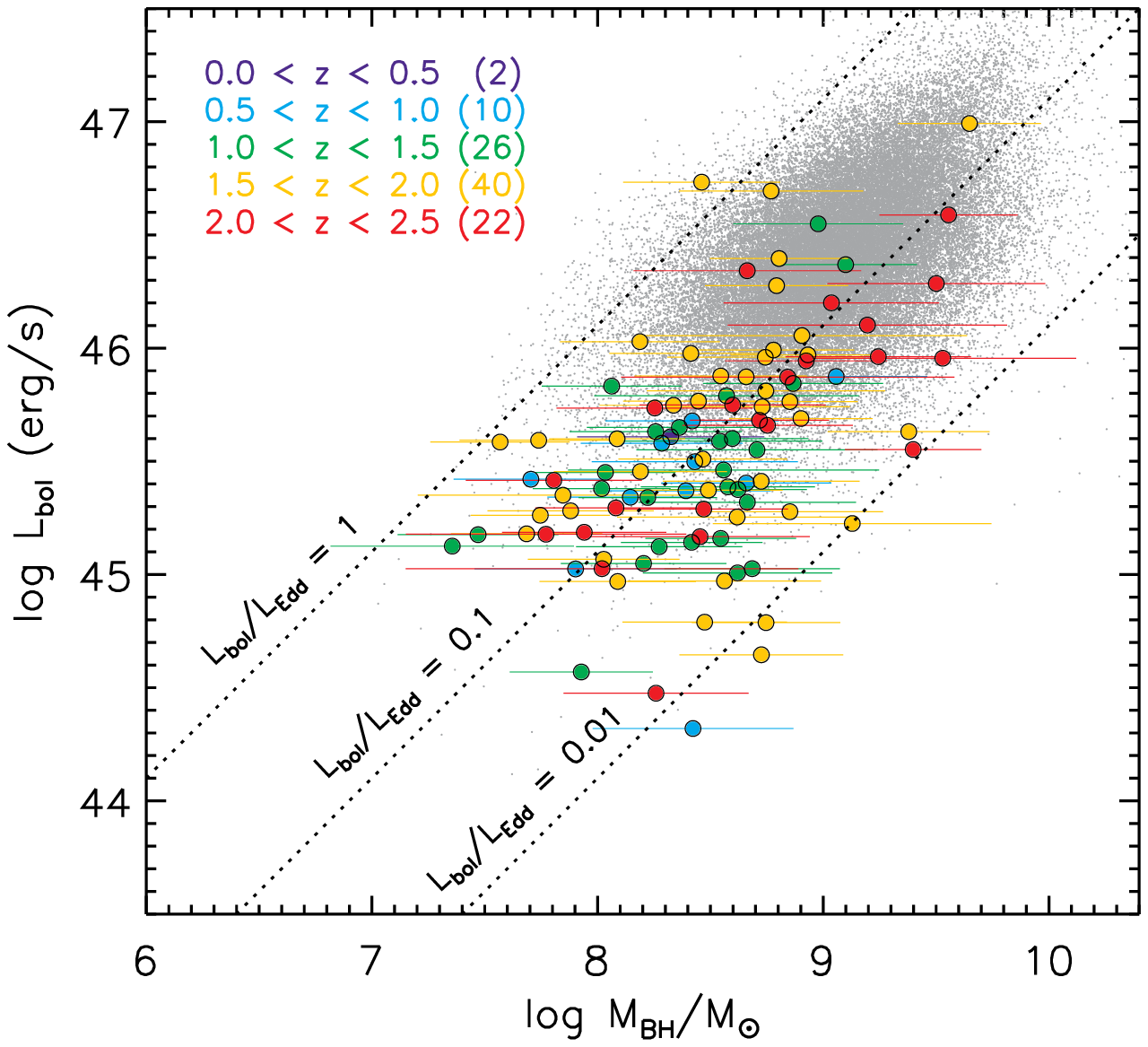}
\caption{AGN bolometric luminosity versus black hole mass for our sample of broad-line AGNs in the different redshift bins. The numbers in parentheses refer to number of sources in each redshift bin. The SDSS quasar sample (gray points; \citealt{Shen11}) is shown for comparison. As a reference, lines of constant Eddington ratio ($L_{\rm bol}/L_{\rm Edd}$) equal to 1, 0.1, and 0.01 are plotted as dotted lines.}
\label{fig:LbolMbh}
\end{figure*}

In short, we apply the absolute flux calibration by matching the deep ground-based photometry available in the COSMOS field. For~\Ha~we fit and subtract a power-law continuum, $f_{\lambda} \propto \lambda^{-\alpha}$, from the spectra, and fit the narrow emission lines of \NII~$\lambda6548,~6583\AA$ with a fixed ratio of 2.96, and \Ha~$\lambda6563\AA$. The broad \Ha~line is fit by one or two broad Gaussian components. In the case of the \Hb~and~\MgII~lines, we simultaneously fit the combination of the power-law continuum and the \FeII~emission components (e.g., \citealt{Vestergaard01, Matsuoka07, Harris13}), for which we adopt an empirical template from \citet{Vestergaard01} convolved with Gaussian profiles of various widths. Then we subtract the best-fit power-law continuum and the \FeII~emission complex from the spectra. For \Hb, the narrow line model is composed of \Hb~$\lambda4861\AA$ and \OIII~$\lambda4959,~5007\AA$ with a fixed ratio of 2.98. We fit the broad \Hb~line with one or two broad Gaussian components. For \MgII, we fit with one or two broad Gaussian components. Finally, we measure the broad-line width and the continuum/line luminosity from the best-fit spectra. The uncertainties of the continuum luminosity are calculated from the average of noise spectrum in the continuum wavelength range. Figure~\ref{fig:Mbh_fits} shows examples of broad-line fits for the \Ha~(top), \Hb~(middle), and \MgII~(bottom) emission lines, respectively.

We compute BH masses using the equation in \citet[\Ha~and~\Hb; provided by \citealt{Vestergaard06}]{Schulze18} and \citet[\MgII]{Trakhtenbrot12}:
\begin{equation*}
M_{\rm BH} = 10^{6.71}(\frac{L_{H\alpha}}{\rm 10^{42}~ergs~s^{-1}})^{0.48}(\frac{\rm FWHM_{H\alpha}}{\rm 10^{3}~km~s^{-1}})^{2.12}~\rm M_{\odot} \\
\end{equation*}
\begin{equation*}
M_{\rm BH} = 10^{6.91}(\frac{L_{5100}}{\rm 10^{42}~ergs~s^{-1}})^{0.50}(\frac{\rm FWHM_{H\beta}}{\rm 10^{3}~\rm km~s^{-1}})^{2.0}~\rm M_{\odot} \\
\end{equation*}
\begin{equation*}
M_{\rm BH} = 10^{6.748}(\frac{L_{3000}}{\rm 10^{44}~ergs~s^{-1}})^{0.62}(\frac{\rm FWHM_{MgII}}{\rm km~s^{-1}})^{2.0}~\rm M_{\odot} \\
\end{equation*}

\noindent where FWHM is the FWHM of the line in units of 1000 ${\rm km~s^{-1}}$, and $L_{H\alpha}$ is the luminosity of broad~\Ha~line. The $L_{5100}$ and $L_{3000}$ are the monochromatic continuum luminosities at rest-frame 5100\AA~and 3000\AA, respectively. The number of sources whose $M_{\rm BH}$ were derived using \Ha,~\Hb,~and~\MgII~lines are 21 (2 from DEIMOS, 19 from FMOS), 5 (2 from DEIMOS, 3 from FMOS), and 74 (all from DEIMOS), respectively. 

The BH estimators we used are based on the mean virial coefficient $\epsilon\sim1$ (e.g., \citealt{Onken04, Grier13}), corresponding to the mean virial factor of $f_{\rm vir}\sim4-5$, which is slightly higher than, but consistent with independently calibrated by \citet[$f_{\rm vir}\sim3.5$]{Grier17} via direct modeling of the BLR structure and dynamics. While the measurement uncertainties on $M_{\rm BH}$ are relatively small ($\sim0.1$ dex), the systematic uncertainties associated with different single-epoch virial calibrations carry a scatter of $\sim0.3$ dex (e.g., \citealt{McGill08, Trakhtenbrot12, Shen13}). We determine the uncertainties of BH mass given by the sum of the statistical and systematic uncertainties.

In Figure~\ref{fig:LbolMbh} we show AGN bolometric luminosity versus BH mass for our sample of broad-line AGNs in the different redshift bins. The AGN bolometric luminosity is derived using the absorption-corrected rest-frame 2-10 keV luminosity by applying a luminosity-dependent bolometric correction described in \citet[see \citealt{Marchesi16, Suh19}]{Marconi04}. For comparison, we show the SDSS DR7 quasar sample (gray points; \citealt{Shen11}), which is limited to the high-mass and high-luminosity regime due to the SDSS spectroscopic follow-up flux limit. Our sample of broad-line AGNs covers the BH mass range $7.0<\log(M_{\rm BH}/M_\odot)<9.5$ and the bolometric luminosity range $44<\log L_{\rm bol}~{\rm erg~s^{-1}}<47$ with the median value of Eddington ratios $L_{\rm bol}/L_{\rm Edd}\sim0.1$. We list our sample of broad-line AGNs in Table~\ref{tbl:sample}, which includes BH masses, AGN bolometric luminosities, and host total stellar masses. 

\section{The $M_{\rm BH}-M_{\rm stellar}$ Relation beyond the local universe}

\begin{figure*}
\centering
\includegraphics[width=0.8\textwidth]{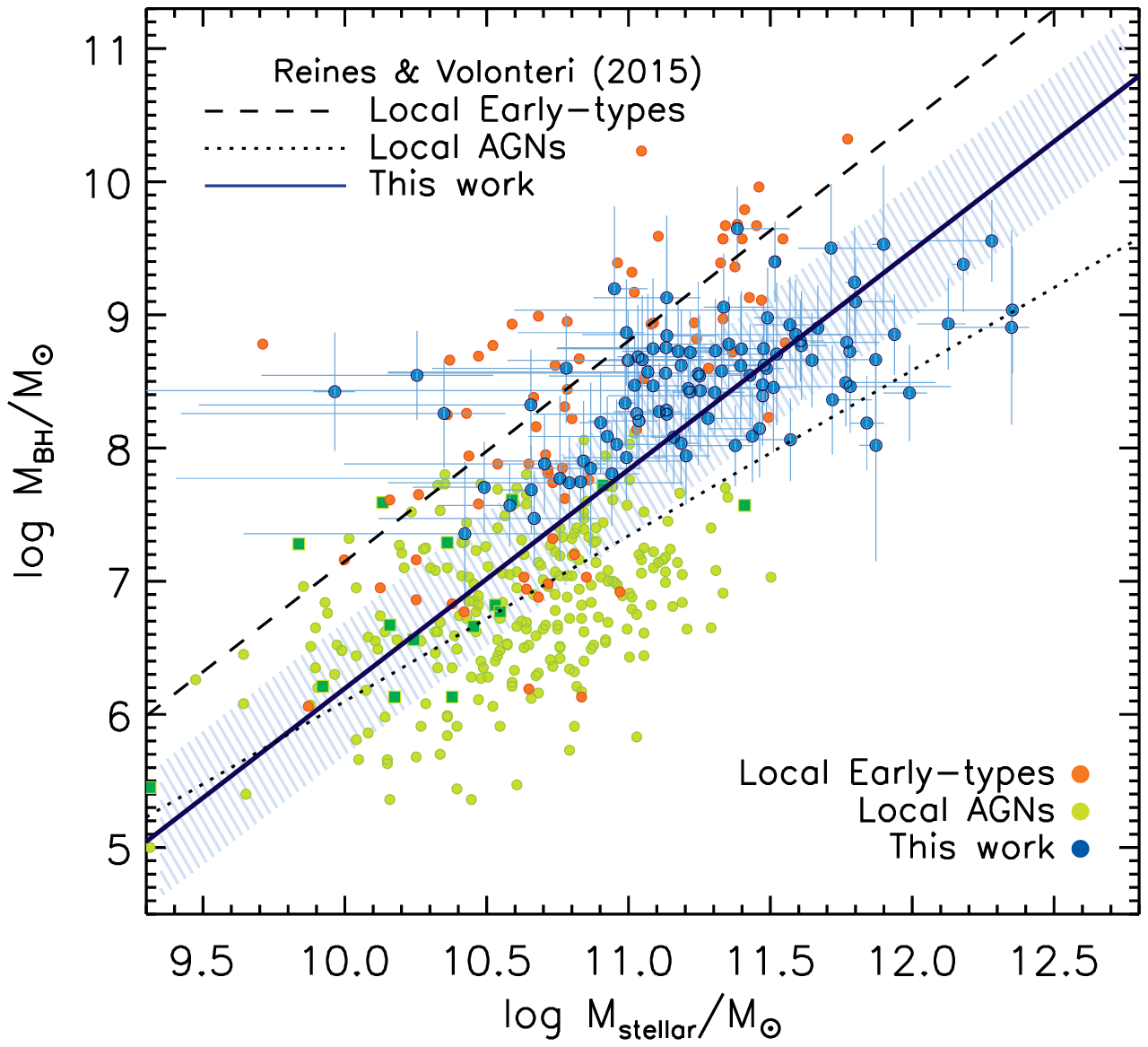}
\caption{BH mass versus galaxy total stellar mass for our sample of 100 broad-line AGNs (blue circles). The orange circles indicate the inactive sample of elliptical galaxies and spiral/S0 galaxies with classical bulges at local universe \citep[taken from \citealt{Kormendy13}]{Reines15}, and the light green symbols indicate the local AGNs from \citet{Reines15}. The 15 reverberation-mapped AGNs from \citet[with BH masses taken from \citealt{Bentz15}]{Reines15} are shown as dark green squares. The dashed and dotted lines indicate the local $M_{\rm BH}-M_{\rm stellar}$ relation derived using total stellar mass from \citet{Reines15}, as a reference. We show the $M_{\rm BH}-M_{\rm stellar}$ relation for the combined sample of local inactive early-types, local AGNs, and our high-z AGNs as a blue solid line with the 1$\sigma$ scatter.}
\label{fig:MbhMs}
\end{figure*}

We present the distribution of our sample of broad-line AGNs and their host galaxies at high redshift on the $M_{\rm BH}-M_{\rm stellar}$ diagram in Figure~\ref{fig:MbhMs}. We also show the inactive sample of early-type galaxies at local universe and the local AGNs taken from \citet{Reines15}\footnote{We correct the stellar masses of \citet{Reines15} sample to match the mass-dependent offset following the procedure outlined in \citet{Shankar19}. \citet{Reines15} adopt BH masses based on a mean virial factor of $f_{\rm vir}=4.3$ from \citet{Grier13}, consistent with our value.}. \citet{Reines15} sample includes 262 broad-line AGNs based on the Sloan Digital Sky Survey (SDSS) Data Release 8 (DR8) spectroscopic catalog, and a subsample of 15 reverberation-mapped AGNs taken from \citet{Bentz15}, as well as 79 galaxies with dynamical BH mass taken from \citet{Kormendy13}.

\begin{figure*}
\centering
\includegraphics[width=1\textwidth]{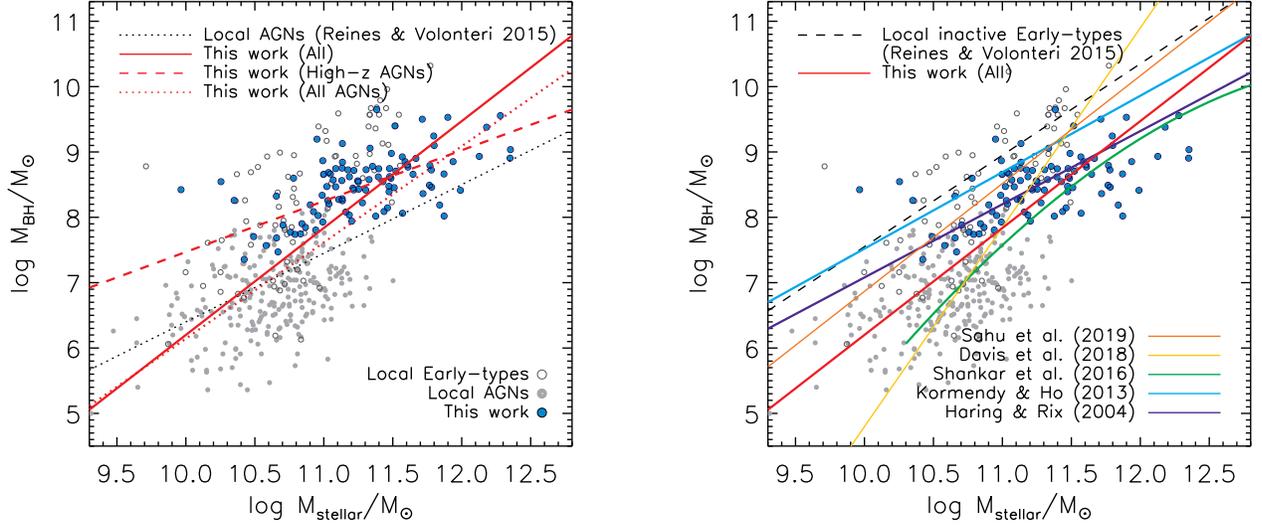}
\caption{Same as Figure~\ref{fig:MbhMs}, but indicated with the local scaling relations for AGNs (left) and those for inactive galaxies hosting dynamically-detected BHs (right), for comparison. The red solid line indicates our derived $M_{\rm BH}-M_{\rm stellar}$ relation for the combined sample of local inactive early-types (open circles), local AGNs (gray circles) from \citet{Reines15}, together with our sample of high-z AGNs (blue circles). The $M_{\rm BH}-M_{\rm stellar}$ relations only for our high-z AGNs and for all combined AGNs (local AGNs + our high-z AGNs) are indicated as a red dashed and dotted lines, respectively. The local $M_{\rm BH}-M_{\rm bulge}$ relations from \citet[purple]{Haring04} and \citet[blue]{Kormendy13} are indicated. The local $M_{\rm BH}-M_{\rm stellar}$ relations for early-type galaxies \citep[orange]{Sahu19} and those for late-type galaxies \citep[yellow]{Davis18} are also plotted. We also show the local intrinsic/unbiased $M_{\rm BH}-M_{\rm stellar}$ relation predicted by \citet{Shankar16} plotted as a green curve.}
\label{fig:MbhMsfit}
\end{figure*}

We derive the linear correlation between BH mass and total stellar mass for the combined sample of local inactive early-types, local AGNs \citep{Reines15}, and our sample of high-redshift AGNs using a Bayesian approach to take into account uncertainties \citep{Kelly07}. We find the relation as
\begin{equation*}
\begin{split}
log~M_{\rm BH}/M_{\odot} = & (1.47\pm0.07)\times~log~M_{\rm stellar}/M_{\odot} \\
 & - (8.56\pm0.04) \\
\end{split}
\end{equation*}
with the intrinsic scatter of $\sim$0.5 dex. We also derive the $M_{\rm BH}-M_{\rm stellar}$ relation for our sample of high-redshift AGNs alone, and for the combined sample of local AGNs and our high-redshift AGN sample. The derived linear relations are shown in the Figure~\ref{fig:MbhMs} and~\ref{fig:MbhMsfit}.

\begin{figure*}
\centering
\includegraphics[width=0.9\textwidth]{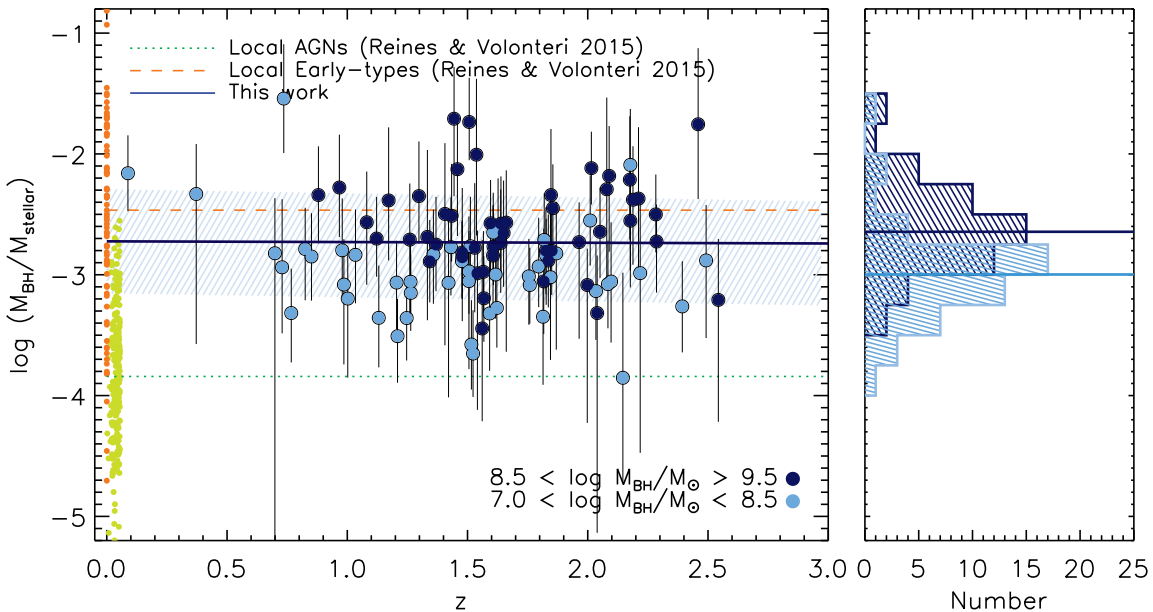}
\caption{Evolution of $M_{\rm BH}$ to $M_{\rm stellar}$ ratios. The green dotted and orange dashed lines represent the median $M_{\rm BH}$/$M_{\rm stellar}$ values of local AGNs (green circles) and those of local inactive early-type galaxies (orange circles) from \citet{Reines15}, respectively. The solid line represents the best-fit trend for our sample of AGNs. We divided our sample into AGNs with less massive BHs ($7.0<\log~M_{\rm BH}/M_{\odot}<8.5$) and AGNs with massive BHs ($8.5<\log~M_{\rm BH}/M_{\odot}<9.5$).}
\label{fig:MbhMs_z}
\end{figure*}

Our sample of AGN host galaxies at high redshift shows a modest offset from the local AGN relation \citep{Reines15}, while the mass range of our sample corresponds to that for the most massive early-type galaxies with inactive BHs in the local universe. It seems that AGN host galaxies beyond the local universe occupy, on average, a region between the relation of local AGNs and those of local early-type galaxies with inactive BHs. We note that our sample of AGNs has a lower limit to the BH mass of $\log M_{\rm BH}\sim7.04 M_{\odot}$ at $z=2$.

We compare our results with the local $M_{\rm BH}-M_{\rm bulge}$ relations in the right panel of Figure~\ref{fig:MbhMsfit} \citep{Haring04, Kormendy13}. Our sample of AGNs tend to lie below the \citet{Kormendy13} relation for local inactive early-type galaxies at a given $M_{\rm stellar}$. It appears that the local $M_{\rm BH}-M_{\rm stellar}$ relation of \citet{Davis18} is consistent with our data, even though only based on late-type galaxies, while the relation of \citet{Sahu19},  which is only based on early-type galaxies, seems to be inconsistent with our high-redshift AGNs. \citet{Shankar16} proposed the local intrinsic relation by correcting the selection bias in the $M_{\rm BH}-M_{\rm stellar}$ relation of inactive BHs with dynamically-measured masses. Indeed, we show that even our high-redshift sample of AGNs well matches the proposed local intrinsic  $M_{\rm BH}-M_{\rm stellar}$ relation within the scatter, further corroborating the presence of a bias in local inactive BHs, and a negligible evolution in the intrinsic  $M_{\rm BH}-M_{\rm stellar}$ relation at least up to $z\sim2.5$. Recently, \citet{Izumi19} also found that the low-luminosity quasars discovered by Subaru High-z Exploration of Low-Luminosity Quasars (SHELLQs) survey appear to be located on or even below the local $M_{\rm BH}-M_{\rm bulge}$ relation, even at $z\sim6$.

\begin{figure*}
\centering
\includegraphics[width=0.8\textwidth]{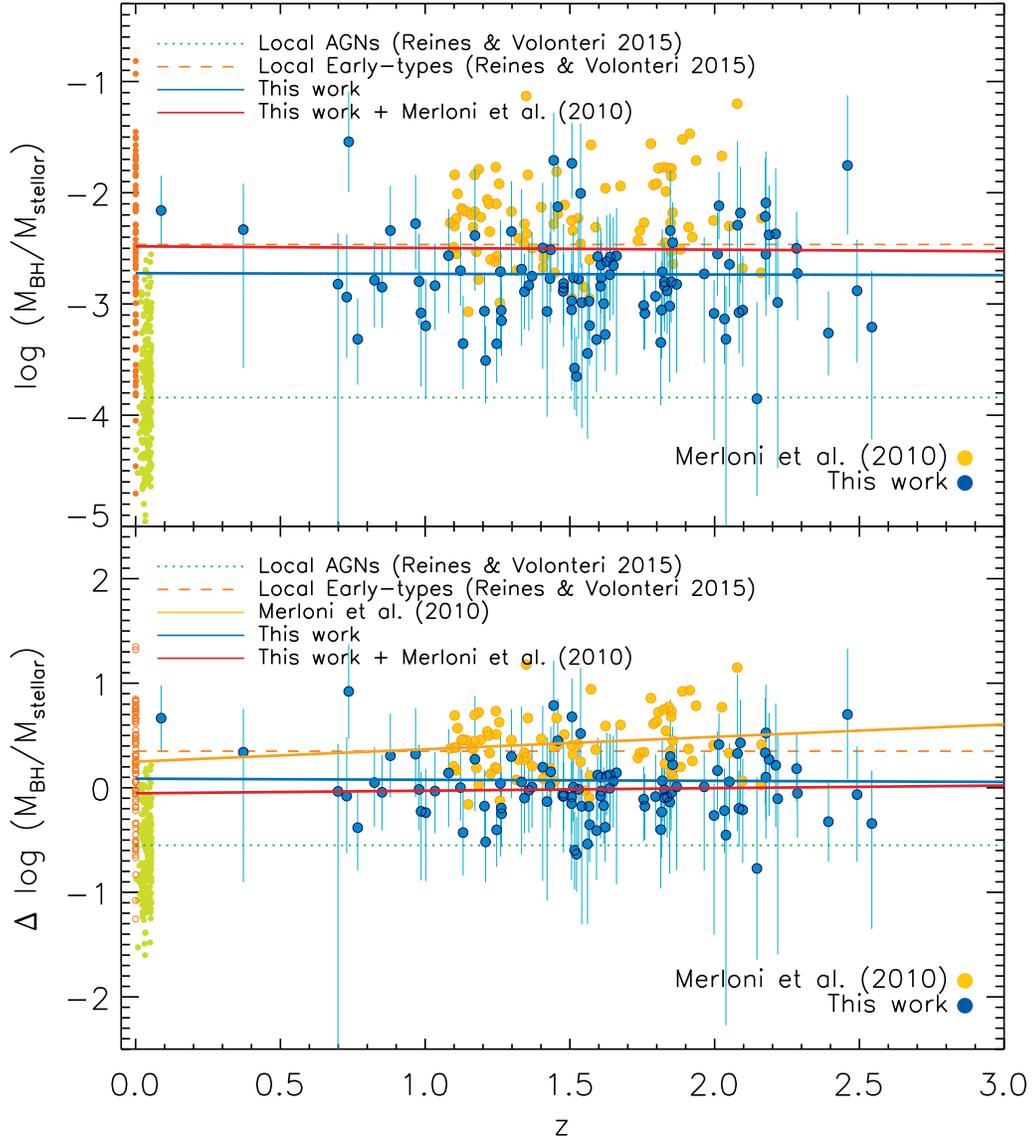}
\caption{Top: redshift evolution of $M_{\rm BH}$-to-$M_{\rm stellar}$ ratios for our sample of AGNs (blue) in comparison with data from the literature, plotted as yellow symbols \citep{Merloni10}. The orange and the green circles indicate the individual local inactive early-type galaxies and the local AGNs from \citet{Reines15}. Blue line indicates the best-fit trend for our sample of AGNs (see Figure~\ref{fig:MbhMs_z}). Red line shows the best-fit of combined sample of our AGNs (blue) with the data from \citet[yellow]{Merloni10}.  The dotted and dashed lines represent the median values of local AGNs (green circles) and those of local inactive early-type galaxies (orange circles) from \citet{Reines15}, respectively. Bottom: evolution of the offset ($\Delta~\log M_{\rm BH}/M_{\rm stellar}$) from the local relation of \citet{Haring04}, which is calculated as the distance of each point to the \citet{Haring04} correlation. Yellow line shows the best-fit of \citet{Merloni10}, and blue line shows the best-fit of our sample of AGNs. Red line indicates the best-fit trend for the combination of our sample of AGNs and those of \citet{Merloni10}.}
\label{fig:MbhMs_z_lit}
\end{figure*}

In Figure~\ref{fig:MbhMs_z}, we show the histogram of BH-to-stellar mass ratios ($\log M_{\rm BH}/M_{\rm stellar}$) and the trend with redshift for our sample of AGN host galaxies in the two BH mass bins. We indicate the median $M_{\rm BH}/M_{\rm stellar}$ values for the local AGNs and the local inactive early-type galaxies from \citet{Reines15} as a reference. Overall, we find no significant evolution of the $M_{\rm BH}-M_{\rm stellar}$ relation with $M_{\rm BH}/M_{\rm stellar}\sim0.2$\% up to $z\sim2.5$, which is markedly higher, by an order of magnitude, than local AGNs ($M_{\rm BH}/M_{\rm stellar}\sim0.015\%$), while smaller than the BH-to-bulge mass ratios derived from inactive early-type galaxies ($M_{\rm BH}/M_{\rm bulge}\sim0.4$\%) in the local universe. We note that the BH mass range of local AGNs from \citet{Reines15} is $5.5<\log~M_{\rm BH}/M_{\odot}<8.0$. It seems that galaxies with less massive BHs tend to have smaller $M_{\rm BH}/M_{\rm stellar}$ ratios than those with more massive BHs. It is also interesting to note that the scatter does not depend much on redshift. 

Our results are consistent with some previous studies (e.g., \citealt{Shields03, Salviander07, Jahnke09, Cisternas11, Shen08, Shen15}) of no evolution in $M_{\rm BH}/M_{\rm stellar}$ ratios with redshift, while in contrast to other studies indicating that BHs predate the growth of their host galaxies (e.g., \citealt{Treu07, Shen08, Woo08, Merloni10, Decarli10, Bennert11}). We compare our findings with data from the zCOSMOS bright spectroscopic survey \citep{Merloni10},\footnote{We corrected the stellar masses of \citet{Merloni10} by subtracting $\sim$0.25 dex, to take into account for the different IMF.} in the top panel of Figure~\ref{fig:MbhMs_z_lit}. Compared to a sample of 89 Type 1 AGNs from \citet{Merloni10}, our sample of AGNs, which are selected by deep {\it Chandra} X-ray, were able to fill in the lower BH mass region by reaching to lower luminosities with the deeper spectroscopy that were missed in earlier luminous sample at high redshift. 

We further explore the deviation of our sample of AGN host galaxies from the local scaling relation by measuring the offset, $\Delta~\log M_{\rm BH}/M_{\rm stellar}$, perpendicular to the \citet{Haring04} relation, following a same approach as \citet{Merloni10}. We note that the local sample of \citet{Haring04} is mostly bulge-dominated (i.e., $M_{\rm bulge}\sim M_{\rm stellar}$). The bottom panel of Figure~\ref{fig:MbhMs_z_lit} shows the $\log M_{\rm BH}/M_{\rm stellar}$ relative to the local relation of \citet{Haring04} as a function of redshift. We show the zCOSMOS AGN sample from \citet{Merloni10} with their best-fit relation: $\Delta\log(M_{\rm BH}/M_{\rm stellar})=0.68\log(1+z)$. We conclude that we do not find a significant cosmic evolution of the $M_{\rm BH}/M_{\rm stellar}$ ratio that our sample of moderate-luminosity AGNs, together with the relatively bright AGN sample of \citet{Merloni10} are indeed broadly consistent with the local $M_{\rm BH}-M_{\rm bulge}$ relation \citep{Haring04} up to $z\sim2.5$, in agreement with no evolution of the scaling relation suggested by previous studies (e.g., \citealt{Shields03, Salviander07, Shen08, Cisternas11, Shen15}). 

While there could be potential systematics and selection biases due to incompleteness in stellar mass and BH mass measurements in the resulting $M_{\rm BH}-M_{\rm stellar}$ relation (i.e., \citealt{Lauer07, Shen15, Shankar16, Shankar19}), i.e., a threshold in luminosity and the detectability of the broad emission line, on average, could lead a bias toward more massive BHs with respect to the galaxy stellar mass, we probe sufficiently low-luminosity AGNs at high redshift, covering the population of less massive BHs.

\section{Discussion}

In this work we provide evidence for a non-evolving $M_{\rm BH}-M_{\rm stellar}$ relation. This single observation has profound implications on several aspects of galaxy evolution. Direct time integration of the total emissivity of AGN, suggests a decrease of the BH mass density of a factor of $\sim3$ up to $z\sim2$, assuming a constant radiative efficiency (see e.g., bottom panels of Figure 4 in \citealt{Shankar13}). A constant $M_{\rm BH}/M_{\rm stellar}$ ratio would then imply a proportionally similar evolution in the integrated stellar mass density of galaxies, which might indeed be the case (e.g., Figure 27 in \citealt{Bernardi10}), though systematics in stellar mass measurements still prevent a robust conclusion in this respect (e.g., \citealt{Bernardi17}). This result of a constancy in the $M_{\rm BH}-M_{\rm stellar}$ relation is also in line with other independent lines of evidence. For example, several studies suggested a weak evolution in the $M_{\rm BH}-\sigma$ relation, either based on Soltan-type arguments or direct observations (e.g., \citealt{Gaskell09, Shankar09, Zhang12, Shen15}). 

A non-evolving $M_{\rm BH}-\sigma$ and $M_{\rm BH}-M_{\rm stellar}$ relation would have significant implications on the overall co-evolution of the BHs and galaxies and on the fundamental plane of early-type galaxies. This would necessarily also imply a weak evolution in the relation of the $\sigma-M_{\rm stellar}$ at least in early-type, bulge-dominant galaxies. If in the central regions one can approximate $M_{\rm BH}\sim M_{\rm stellar}\sim M_{\rm dyn}\sim k\times R_{e}/\sigma^{2}$, at fixed BH mass and velocity dispersion one would naively expect a constant product $k\times R_{e}$. If early-type galaxies decrease in size $R_{e}$ at higher redshifts (e.g., \citealt{vanDokkum13}), this would imply a proportional increase in the virial constant $k$, which is expected if the S\'ersic index decreases at earlier epochs (e.g., Figure 1 in \citealt{Bernardi18}), which is expected in merger-dominated models (e.g., \citealt{Hilz13}).

We suggest that the BHs and galaxies both have grown predominantly at higher redshift ($z>3$), and all necessary stellar mass may already exist in galaxies at $z\sim3$. This is in line with the fact that their growth rates seem to be broadly correlated, both in the integrated sense, i.e., SFR density and BH accretion rate density, and the individual sense, i.e., SFR and $L_{\rm AGN}$ (e.g., \citealt{Shankar13, Suh19}). In the later stage of AGN evolution, the moderate-luminosity AGNs with relatively high accreting BHs found below the local scaling relation will presumably move up to the $M_{\rm BH}-M_{\rm stellar}$ plane induced by secular processes, becoming eventually inactive galaxies harboring SMBHs. The secular process may not change the overall $M_{\rm BH}-M_{\rm stellar}$ ratio by feeding both BH and stellar masses. 

We note here that there could be an evolution in the $M_{\rm BH}-M_{\rm bulge}$ relation, considering that our sample of AGNs are not purely bulge-dominated (see e.g., \citealt{Jahnke09}). To explain the tight relationship between the final BH and spheroid bulge mass, a re-distribution of stellar masses is required from the disk to the bulge, also induced by secular processes such as disk instabilities. \citet{Elbaz18} found that the galaxies hosting an AGN at $z\sim2$ appear to be systematically associated with the most compact star-forming galaxies, suggesting that the physical mechanism responsible for the rise in star-formation compactness also efficiently feeds the central BH, or possibly that the AGN activity plays a role in triggering the compact star-formation through positive feedback, i.e., bulge-formation (see also \citealt{Chang17}). These results suggest that the majority of AGN host galaxies at $z<3$ might be driven more by internal secular processes, implying that they have substantially grown at a much earlier epoch. 

\section{Conclusions}

We have studied the evolution of relations between BH mass and total stellar mass for 100 X-ray-selected moderate-luminosity, broad-line AGNs in the CCLS up to $z\sim2.5$. By taking advantage of the deep multi-wavelength photometry and the unique, intensive spectroscopy in the COSMOS field, we measured in a uniform way the galaxy total stellar mass using a SED decomposition technique and the BH masses based on single-epoch virial estimates. The main results are summarized as follows.

\begin{itemize}
 \item Our sample of AGN host galaxies has total stellar masses in the range $10^{10-12}M_{\odot}$, and the BH masses are broadly distributed around $10^{7.0-9.5}M_{\odot}$, probing sufficiently low-luminosity AGNs at high redshift covering the less massive BH populations. The median Eddington ratios is $L_{\rm bol}/L_{\rm Edd}\sim0.1$. 
 \item Our sample of AGN host galaxies beyond the local universe tend to lie below the $M_{\rm BH}-M_{\rm stellar}$ relations for local inactive early-type galaxies at a given $M_{\rm stellar}$, with no evident evolution with redshift.
 \item By combining our moderate-luminosity AGNs with the relatively bright zCOSMOS AGN sample \citep{Merloni10}, we find that the distribution of $M_{\rm BH}-M_{\rm stellar}$ relations for high-redshift AGNs is broadly consistent with the local value within uncertainties, with $M_{\rm BH}/M_{\rm stellar}\sim0.3$\% up to $z\sim2.5$.
 \end{itemize}

\acknowledgments

We thank the anonymous referee for valuable comments that significantly improved this work.
FS acknowledges partial support from a Leverhulme Trust Research Fellowship. VA acknowledges funding from the European Union's Horizon 2020 research and innovation programme under grant agreement No749348.


\end{document}